\documentclass[twocolumn,prl,aps]{revtex4}
\usepackage{graphicx}

\begin{document}
\title{Loss of quantum coherence from discrete quantum gravity}
\author{Rodolfo Gambini$^{1}$, Rafael A. Porto$^{1}$ and Jorge Pullin$^{2}$}
\affiliation {1. Instituto de F\'{\i}sica, Facultad de Ciencias,
Igu\'a 4225, esq. Mataojo, Montevideo, Uruguay. \\ 2. Department
of Physics and Astronomy, Louisiana State University,\\ 
202
Nicholson Hall, Baton Rouge, LA 70803-4001}
\date{February 25th 2004}

\begin{abstract}
We show that a recent proposal for the quantization of gravity based
on discrete space-time implies a modification of standard quantum
mechanics that naturally leads to a loss of coherence in  quantum states of the
type discussed by Milburn.  The proposal overcomes the energy
conservation problem of previously proposed decoherence mechanisms
stemming from quantum gravity. Mesoscopic quantum systems (as
Bose--Einstein condensates) appear as the most promising testing
grounds for an experimental verification of the mechanism.
\end{abstract}

\maketitle

It is commonly believed that a satisfactory theory of quantum gravity
may require a drastic modification of our description of
space-time. Among the proposed fundamental changes that appear in the
literature is that the ultimate theory may imply a discrete structure
for space-time at a microscopic level. Recently, a
quantization of gravity in discrete space-time has been developed
which addresses major fundamental concerns of the canonical program
\cite{discrete}.  Among the appealing elements of the proposal is the
solution of the problem of time in generally covariant systems through
the introduction of relational time in quantum mechanics
\cite{greece}. Having promoted time into the quantum realm it is
meaningful to ask how to compare the resulting theory with the
traditional Schr\"odinger picture. We shall show in what follows that
in the semiclassical limit, discrete quantum gravity may lead to
information loss in quantum states. We will also argue that the
process can potentially lead to observable consequences. This
construction leads to the same density matrix evolution equation that
has been considered by several authors \cite{lind,ghir,mil,perc,adler}
in other contexts. It can also be viewed as a concrete implementation
of the idea of Penrose
\cite {penrose} that gravity should play a role in the state reduction
process.  The equation is given by,
\begin{equation}
\frac{\partial\rho}{\partial t}=-i[H,\rho]-{\cal
D}(\rho)\label{rho},
\end{equation}
with ${\cal D}(\rho)$ a decoherence term which has been usually
taken as having the modified Lindblad's form\cite{lind},
\begin{equation}
{\cal D}(\rho)=\sum_n[D_n,[D_n,\rho]], \;\;\;
D_n=D_n^{\dagger},\;\;\; [D_n,H]=0,
\end{equation}
so it defines a completely positive map on $\rho$, and which is
consistent with the monotonous increase of Von Neumann entropy $S={\rm
Tr}(\rho \log \rho)$ and conservation of energy. This type of equation
was introduced by Ghirardi, Rimini and Weber (GRW) \cite{ghir} with
the aim of providing an objective solution to the measurement problem
in standard quantum mechanics. (Similar equations can be used to
describe the decoherence due to interaction with an environment, see
\cite{joos}.) GRW considered a single $D$ as a localizing operator in
coordinate space. As discussed by Adler and Horwitz \cite{adler}, and
also Milburn, Percival and Hughston
\cite{mil,perc}, setting $D$ to be proportional to 
$H$ is most natural since it leads to an objective state vector
reduction in the energy pointer basis. This loss of coherence may be a
way to avoid macroscopic superpositions, like the ``Schr\"odinger
cat'' \cite{mil,adler}.  We shall show that starting from the recent
proposal for ``consistent discrete quantum gravity'' of
\cite{discrete} one obtains proportionality between $D$ and $H$. There
are other type of constructions where ${\cal D}$ represents
collectively quantum gravity effects. In general it does not take the
commutator form of equation (2) and it is mainly associated with
partial traces over ``environmental" degrees of freedom like
microscopical black holes or strings \cite{haw,ellis,banks}. As shown
by Hawking \cite{hawk}, these generalized equations may lead to
violations of unitarity during black hole evaporation. Hawking's
proposal was criticized on the grounds that it violates energy
conservation \cite{banks}. To find an explicit description of
information loss in quantum gravity consistent with energy
conservation is one of the major challenges of the field.  We will
show that the ``consistent discrete approach'' \cite{discrete} ensures
energy conservation. Therefore one can conjecture that this
description may provide a concrete theoretical solution to the black
hole entropy problem based on a quantum description of space-time.

Let us now show how discrete quantum gravity leads to $D \propto H$. There
have been several attempts to formulate a purely relational
description of quantum mechanics \cite{relational,page}. The idea is
to consider a system that is closed and try to define a notion of time
through the evolution of a sub-system of the closed system. This is of
interest conceptually in ordinary quantum mechanics, since it frees
the description of its reliance on an external classical clock and is
unavoidable when one is considering quantum cosmology, where there
simply does not exist an external observer. Perhaps the most explicit
examples of attempting to present a detailed description at the
quantum mechanical level of a purely relational evolution are the
series of papers by Page, Wootters and others \cite{page}.  A notion
of time is introduced via conditional probabilities, that is, asking
what is the probability that a certain quantity has a certain value
when another quantity has a given value. The latter could be viewed as
a ``quantum mechanical clock'' introducing a purely quantum mechanical
notion of time. The resulting description is an extension of ordinary
quantum mechanics that allows to make predictions in regimes where the
notion of a classical clock is not applicable. As discussed by
Kucha\v{r} \cite{Kuchar}, there are problems with the relational
approach, when applied to totally constrained systems like general
relativity due to the presence of constraints. The proposed consistent
discrete framework for gravity is constraint-free, and therefore we
have been able to show that it is possible to introduce a relational
description of time avoiding the hard issues that plagued the
previous attempts \cite{greece}.

Let us consider a concrete application of the consistent discrete
approach to a Friedmann cosmology coupled to a scalar field. We
will denote the latter by $\phi$. Details of a similar model can be seen
in \cite{cosmo}. In a Friedmann cosmology set up as a canonical
system, the fundamental variable is the spatial metric, which has
only one independent component and its canonically conjugate
momentum.  It is best to describe the model using Ashtekar
variables \cite{Ashtekar}. In these variables the metric is
replaced by a triad, which also has only one independent component
in the Friedmann case. We denote it by $E$. Its canonically
conjugate momentum is denoted by $A$. The scalar field and their
canonical momenta are $(\phi,P^\phi)$.  One writes the Einstein
action for the model and discretizes the evolution parameter
(since the model is homogeneous) as $\tau=n\Delta \tau$. Evolution
can be represented at the canonical level via a canonical
transformation that implements the discrete equations of motion
from the level $n$ to level $n+1$. One of the equations (the one
that would correspond in the discrete theory to the single
constraint of the continuum theory) determines the lapse and is
solved for it. The resulting theory therefore has no constraints.
All its variables are therefore candidates for physical
observables and one is therefore ready for the application of
the relational time formalism. The quantization of the theory is
given in terms of quantum states $\Psi(A,P^\phi,n)$. The evolution
of the theory is unitary in terms of the $n$ variable, since we
can represent the canonical transformation as a unitary evolution
operator in the quantum theory. We can now use for instance the
variable $A$ as a time variable and compute the conditional
probability that the scalar field momentum have a certain value at
a given ``time'' as (one should really phrase it in terms of
intervals since the variables have continuous spectrum)
\begin{eqnarray}
&P_{\rm cond}(P^\phi=x| A=t)=\nonumber&\\&\lim_{N\to \infty}
{\sum_{n=0}^N\Psi^2[A=t,P^\phi=x,n] \over \sum_{n=0}^N 
\int_{-\infty}^{\infty}
\Psi^2[A=t,P^\phi,n] dP^\phi}&
\end{eqnarray}

In the discrete approach the cosmology appears as a succession of
``snapshots'' labeled by the integer $n$, which lacks any
intrinsic meaning. The emergence of time in the model is only
through the correlations of the dynamical variables of the theory.
The quantum theory that results from the relational probabilities
only agrees with ordinary quantum mechanics in regimes in which a
notion of classical time is a good approximation to the behavior
of a quantum variable \cite{greece}. It is clear that generically
there could be departures from this regime. Let us denote by $\rho$
the initial density matrix for the gravity-matter system and
assume that in the semiclassical limit we can decouple the system
as $\rho \approx \rho_1\bigotimes \rho_2$, with $\rho_1,\rho_2$,
associated to geometry and the field respectively. 
We sketch a proof  that there exists a
relational Schr\"odinger picture where there is an effective
density matrix for the ``system" which evolves in ``internal
clock" time into a statistical mixture even if it was in a pure
initial state with a resulting evolution equation of the form (1,2)
with $D$ proportional to $H$.

We discuss the derivation for a generic system, to particularize it to
the cosmology we mentioned before one chooses one of the variables as
clock, for instance $t=A$ and $x=P^\phi$. We proceed in two
steps. First we introduce the relational Schr\"odinger picture that
our approach follows. We compute the evolution of the density matrix
for the ``rest'' system by summing over all configurations in the
variable $n$ that are compatible with a certain time $t$,
$\rho_2(t)\equiv \sum_n {\cal P}_{n}(t)
U_2(n)\rho_2(0){U_2}^{\dagger}(n)$, where ${\cal P}_{n}(t)= {\rm Tr}[
P_{t}(0)U_1(n)\rho_1 U_1(n)]$. Where $U_1(n),U_2(n)$ are the unitary
discrete evolution operators in terms of the discrete parameter $n$
for the ``time'' and ``the rest" variables respectively. We also
denote as $P_t(0), P_x(0)$ the projectors in the $n=0$ level into the
sub-space corresponding to the values $t,x$ of the ``time" and the
field. We will assume now that there exists a Hamiltonian operator for
$U_2$ such that $U_2(n)=\exp(-iH_2n)$. It is important to notice that
this evolution differs from the usual Schr\"odinger picture due to the
presence of the sum, which is related to the fact that there is not a
unique correspondence between the discrete parameter $n$ and a given
value of time $t$. Therefore the trace of the square of the evolved
density matrix will not be one and therefore a pure state evolves into
a mixed state. For reasons of space, we do not present the detailed
derivation that this evolution equation implies the equations (1,2),
but we give the following intuitive explanation. If the probability
distribution ${\cal P}_{n}(t)$ were a Dirac delta (one step $n$ is
associated uniquely to some time $t$) then the density matrix would
satisfy equation (1) with ${\cal D}=0$.  In practice, the probability
distribution will be peaked around some value of $n$ with
contributions from neighboring values. In the sum this implies that
there will be terms representing the evolution from those neighboring
values of n. This evolution can be viewed as generated by the action
of a Hamiltonian operator. This additional operatorial action is what
leads to the double commutator in (2). This can be worked out in
detail, we will present the calculation elsewhere.

Let us make, however, some comments. Consider $t_{max}(n)$, that is,
the value of the maximum probability for the variable $t$ as a
function of $n$. We will assume we chose the temporal variable in such
a way that $t_{max}(n)=\gamma  n$ with $\gamma$ a
constant of the motion and we will denote by $\delta\gamma$ its quantum fluctuations.
It is possible to show that $D=\sqrt{\sigma} H$ with $\gamma \sigma =
\partial (\delta t_{max})^2 /\partial n$, where $\delta
t_{max}=(\delta \gamma) n$. What is happening here is that one chose a
wavepacket in which $t$ was a peaked function of $n$ for the clock,
but as the system evolves such wavepacket spreads out and $\sigma$
will be a measure of that spread. $\sigma$ is related to the rate
of growth of the spread of the packet.

Equations (1,2) imply that coherence is lost since the off diagonal
terms of the density matrix go to zero,
\begin{equation}
{\rho_2}_{nm}(t)= {\rho_2}_{nm}(0)e^{-i\omega_{nm}t}
e^{(-{\sigma}(\omega_{nm})^2)t}\label{sol}
\end{equation}
where $\omega_{mn}=E_m-E_n$ are the Bohr frequencies (for a 
derivation see \cite{Garay}, where this formula was obtained
by studying a classical non-ideal clock). Notice the
loss of coherence implied by the exponential. 
This can have remarkable effects, for
instance if one waits long enough all off-diagonal elements of the
density matrix vanish. In spite of this, it is difficult to find
experimental situations where these effects are measurable. For
instance one may consider light that propagates from distant stars in
order to have long times of flight and enhance the effect. However,
since most optical measurements imply measuring second (or, more
generally, even)-order correlations \cite{glauber}, this loss of
coherence has no visible consequences in optics.

In order to seek for possible experiments, let us first estimate the
value of $\sigma$, which we expect is going to be a small time scale,
of the order of the Planck one. We shall begin by using the whole
universe as the quantum clock in order to get some intuition to the
bounds for these effects to happen. Since the universe is the biggest
reservoir we have it is naturally to believe that it is the best clock
we can build. Let us assume that our present universe may be
modeled by the Friedmann cosmology we discussed before.  As it is
shown in \cite{discrete} the discrete evolution for the connection $A$
goes as $l_p \Lambda^{1/2} a(n+k)^{2/3}$ and
$E={l_p}^2a^2(n+k)^{4/3}$, where $a>0,k$ are two non-dimensional
constants which parameterize the set of equivalent orbits in the
continuum limit. Let us now take the relational time as
$t={\Lambda}^{-3/4}l_p^{-1/2} A^{3/2}$, such that it has dimension of time, and
time is thus measured in cosmological units (we also have
$\hbar=c=1$). We have $t= {l_p} a^{3/2}(n+k)$ which
is of the form $t=\gamma(n+k)$, with
$\gamma={l_p}a^{3/2}$, and linearly in $n$ as we
wanted. On the other hand, due the uncertainty
principle for $E,A$, i.e. $\Delta E \Delta A
> {l_p}^2$, we have $ a {\delta a}^2 > {1\over
l_p\Lambda^{1/2} (n+k)}$. From this one can immediately
estimate a lower bound for $\delta \gamma$ and therefore
 a lower bound for $\sigma>l_p/(t\sqrt{\Lambda})$. For the present
epoch of the universe one therefore has $\sigma>l_p$.
These estimations for $\sigma$
should not be seen as a concrete calculation. In order to refine them
we need to construct a more realistic model of the universe including
more complexity. See \cite{GaPoPu} for a lengthier discussion.

As was first discussed by Ellis et al. \cite{ellis} there are several
possible phenomenological implications of this non standard quantum
behavior. Examples are neutron interferometry and the neutral kaon
decay.  It is easy to see that our approach implies extremely small
corrections to the usual quantum mechanical predictions for these
systems.  All these models suffer from the same problem, they involve
small energy differences between channels so our predictions, though of
theoretical interest, are extremely small. This contrasts with the
string theory predictions considered in \cite{ellis} where the size of
the effect is controlled by the external string background and its
energy scale is conjectured to be some exponent of the Planck
Mass \cite{banks}.

A system which has been considered in connection with Milburn's
\cite{mil} type of decoherence is a two level atom interacting with
the electromagnetic field in a cavity \cite{subir}. We follow closely
the analysis of reference \cite{subir}. The system is described by the
Hamiltonian $H=H_a+H_f+H_i$ where $H_a={\omega \over 2} R_3$ describes
the energy splitting of the two level atom. $H_f=\omega a^{\dagger}a$
is the traditional number operator for the $\omega$-mode of the field
inside the cavity, and $H_i=\lambda(R^{+}a+a^{\dagger}R^{-})$ with
$\lambda^2\sim \mu^2\omega$, the dipole coupling constant. Here $\mu$
is the dipole matrix element between both levels of the atom. The
operators $R_3,R^{\pm}$ are essentially the traditional angular
momentum operators and $(R_3)^2=1$. Notice that this model could be
used to describe any two level system coupled to a mode of a field
which induces transitions. Indeed the population of the upper level is
given by $1/2(1+<R_3>)$.  It is easy to see now that within our
approach the atomic inversion evolves as,
\begin{equation}
<R_3>_{\sigma}=\sum_s
|Q_s(0)|^2e^{\left(-2(s+1)\sigma\lambda^2t\right)}
\cos(2\sqrt{s+1}\lambda t)\label{r3}
\end{equation}
where $Q_s\equiv e^{-|\alpha|^2/2}\frac{\alpha^s}{\sqrt{s!}}$.  We
have considered initially the field in a coherent state $|\alpha>$
($\bar n=|\alpha|^2$), and the atom in its excited state. This system
was first analyzed by Moya et al. \cite{moya} who studied the same
model within Milburn's proposal\cite{mil} which is dynamically
identical to our approach plus the identification $\sigma \rightarrow
\theta_0$.  The physical conclusions are that coherence is destroyed
and ``revivals'' in the coherence are exponentially suppressed. The
loss of coherence is likely to act slower than the
decoherence due to ambiental factors and therefore one expects that
actual experiments will be dominated by the ambient decoherence.  In
fact, modeling the latter by the introduction of a damping constant
$\gamma$ and Langevin noise leads to a solution for the atomic
inversion in the under-damping limit $\gamma << \lambda$, of the
form \cite{subir},
\begin{eqnarray}
<R_3>_{\gamma}=\frac{1-2{\bar n}}{1+2{\bar n}}
\left(e^{-\gamma\frac{1+2{\bar n}}{1+{\bar n}
}t}-1\right)+\nonumber\\ \sum_s |Q_s(0)|^2e^{-\gamma
t}cos(2\sqrt{s+1}\lambda t)
\end{eqnarray}
Notice that both effects are similar but leave a different imprint
since the exponential decay does not depends on the number of photons
$n$ in the ambient decoherence while it does in the one due to loss of
coherence. In order for the effect to be visible one would need a high
intensity laser, which would face limitations since high intensities
will increase the ambient decoherence of the cavity. It should be
noticed that the relative importance of the loss of coherence
can be enhanced through the increase of the dipole
coupling of the field with the two level atom $\mu$. Therefore an
optimal experiment should have intense fields, strong coupling of the
atom to the field and widely separated energy levels in the atom. At
present this appears beyond the state of the experimental art with
these types of experiment.

Since the effect depends on the number of photons, this suggests that
systems involving mesoscopic quantum states should be well suited for
experimentally probing the effect. An example of such systems could be
the Bose-Einstein condensation (BEC).  Recently Greiner at al
\cite{nature} studied the ``collapse and revival'' of matter wave fields
in BEC. They showed that the macroscopical wave function undergoes a
series of ``collapses and revivals'' due the collisions of cold atoms
confined to a potential well. We will not repeat the calculation
explicitly here for this model for reasons of space, but we have found
an effect very similar to that of equation (\ref{r3}) for this
system. The main difference is that in the exponent the number of
atoms enters quadratically, as opposed to (\ref{r3}) where the number
of photons entered linearly. Although the currently considered
experiments with BEC involve very few atoms, and again one is very far
away from seeing quantum gravitational effects, it is possible that
future experiments with more atoms, giving the quadratic dependence on
the number could be more promising. To our knowledge these detection
of revivals will, in the future, provide the most stringent bounds on
quantum gravity inspired decoherence known up to now. In general
experiments that test ``Schr\"odinger cat'' type situations can all
potentially lead to observations of the decoherence we found.  For a
recent review see \cite{Leggett}. SQUID experiments seem to provide
the best bound up to date \cite{Ellis,Joychristian}.

Summarizing, we have shown that a recently introduced proposal for
quantizing gravity on discrete space-time leads naturally to a quantum
mechanics that includes a fundamental decoherence of the Milburn
type. Contrary to Hawking's information loss proposal, ours does not
violate energy conservation. Future experimental developments in
quantum mesoscopic systems can lead to a confirmation of
the process.

We wish to thank Peter Knight for correspondence. This work was
supported by grant NSF-PHY0244335 and funds from the Horace Hearne
Jr. Institute for Theoretical Physics.

\end{document}